# Obfuscation in Bitcoin: Techniques and Politics[1]

Arvind Narayanan & Malte Möser
Princeton University

**Abstract.** In the cryptographic currency Bitcoin, all transactions are recorded in the blockchain — a public, global, and immutable ledger. Because transactions are public, Bitcoin and its users employ obfuscation to maintain a degree of financial privacy. Critically, and in contrast to typical uses of obfuscation, in Bitcoin obfuscation is not aimed against the system designer but is instead enabled by design. We map sixteen proposed privacy-preserving techniques for Bitcoin on an obfuscation-vs.-cryptography axis, and find that those that are used in practice tend toward obfuscation. We argue that this has led to a balance between privacy and regulatory acceptance.

**Obfuscation Techniques**

Bitcoin's design is centered around a widely distributed, global database which stores all transactions that have ever taken place in the system. Thus, there is no avenue for redress if a user wishes to retrospectively hide a transaction. Further, nothing in the ledger is encrypted, and digital signatures are mandatory, ensuring cryptographic attribution of activities to users. On the other hand, account identifiers in Bitcoin take the form of cryptographic public keys, which are pseudonymous. Anyone can use Bitcoin "wallet" software to trivially generate a new public key and use it as a pseudonym to send or receive payments without registering or providing personal information. However, pseudonymity alone provides little privacy, and there are many ways in which identities could be linked to these pseudonyms (Narayanan et al., 2016).

To counter this, Bitcoin and its users employ a variety of obfuscation techniques to increase their financial privacy. We visualize a representative selection of these techniques in Figure 1 based on their time of invention/creation and our assessment of their similarity to obfuscation vs cryptography. We make several observations. First, techniques used in Bitcoin predominantly fall into obfuscation, with stronger techniques being used exclusively in alternative cryptocurrencies (altcoins). Second, there is a trend towards stronger techniques over time, perhaps due to a growing interest in privacy and to the greater difficulty of developing cryptographic techniques. Third, obfuscation techniques proposed at later points in time are seeing less adoption, arguably a result of their increased complexity and need for coordination among participants (Möser & Böhme 2017).

---

[1] Presented at the *International Workshop on Obfuscation: Science, Technology, and Theory*, New York University, April 7-8, 2017.

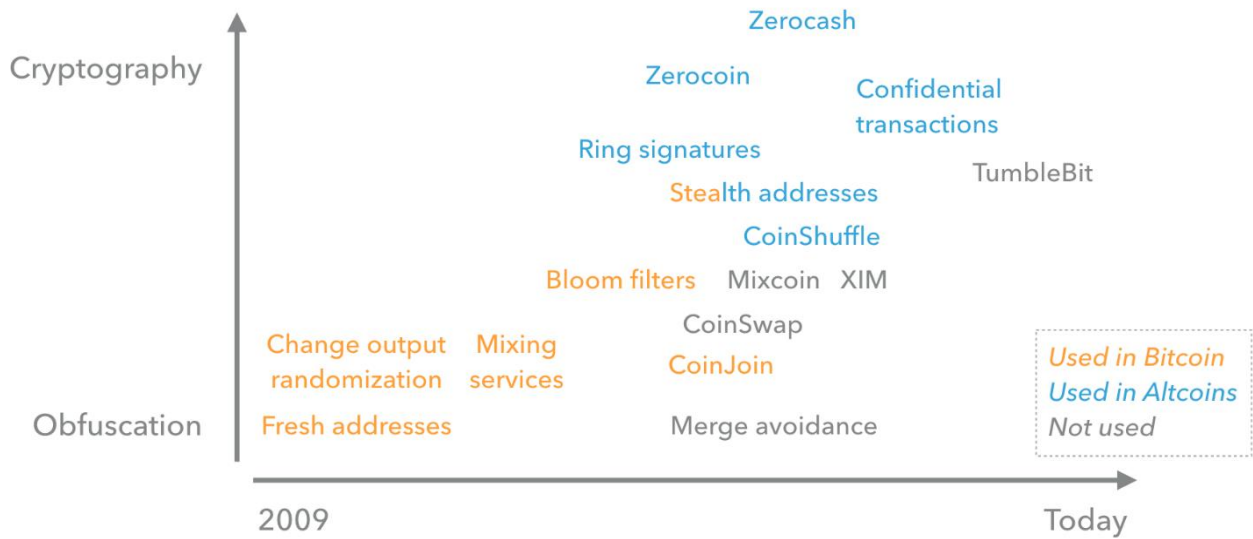

**Figure 1:** Privacy-Enhancing Technologies for Bitcoin. The X-axis is the date of invention and the Y-axis is an informal measure that combines the sophistication of the technique and the strength of the privacy guarantee. See Appendix 1 for references and dates.[2]

Among the techniques used in Bitcoin, the most prevalent can be characterized as "ambiguating obfuscation" (Brunton & Nissenbaum 2015): effectively reducing the information an adversary is able to extract from a particular transaction. Examples include using a new pseudonym for every new transaction and randomizing the structure of transactions to make the spend to the "true" recipient indistinguishable from "change" going back to the sender.

A second type of obfuscation, namely "cooperative obfuscation", has risen in popularity over the last years. For example, users can send their money to a service that will "mix" their funds with those of other users, thereby obfuscating the flow of payments (cf. Möser, Böhme & Breuker 2013). A similar technique called CoinJoin works in a peer-to-peer fashion and doesn't require a trusted intermediary is CoinJoin. Due to the need for these users to find and transact with each other, markets for anonymity have arisen that bring together providers and receivers of anonymity (Möser & Böhme 2016).

**The Case for Obfuscation**

Critically, none of the techniques discussed provide provable privacy guarantees through cryptography. While these do exist and have been deployed (e.g., Zcash), they are far from being adopted by the Bitcoin community, for both technical and political reasons. On the technical side, Bitcoin's decentralization already incurs a severe performance penalty compared to centralized payment systems such as Paypal. Achieving cryptographic privacy would further degrade performance. Obfuscation also has a lighter impact on the usefulness of the blockchain for non-currency applications. The current design allows selectively employing obfuscation, leaving room for other uses that prioritize different goals, such as Colored Coins (Rosenfeld 2012), a protocol for representing assets on top of the Bitcoin blockchain.

---

[2] In a previous draft of this paper, the X-positions of some of the techniques in the figure were slightly off due to an image editing error. We have fixed those, and report the dates in Appendix 1.

On the political side, providing stronger privacy through cryptography might make Bitcoin even more attractive for activities such as money laundering, ransomware, or terrorism financing, and thereby tempt a government crackdown. Much of the Bitcoin community is invested in its mainstream adoption, and therefore keen to avoid such an outcome. When Bitcoin began to be noticed by the press, members of the community went to work explaining it to policy makers. They framed the technology as neutral and unthreatening, and the Bitcoin ecosystem as subject to existing regulations and amenable to new ones (cf. Brito 2013, Brito & Castillo 2013, Lee 2013, Murck 2013, Hattem 2014).

The use of obfuscation in Bitcoin may have achieved a balancing act between the financial privacy of its users and the investigatory needs of law enforcement and regulators. Law enforcement agencies have two important advantages over everyday adversaries: the budget for specialized Bitcoin tracking tools and services (Cox 2017), and subpoena power. The latter allows deanonymizing selected actors by obtaining user records from exchanges and cross-referencing them with the results of blockchain analysis (Meiklejohn et al. 2013). Since only a few governmental actors possess these powers, users still enjoy a measure of financial privacy. Thus, the imperfect privacy protection in Bitcoin may be one of the keys to its success.

**Acknowledgment.** This work was supported by NSF awards CNS-1421689 and CNS-1651938.

**Appendix 1: References for Privacy-Enhancing Techniques for Cryptocurrencies**

| Name | Reference | Approx. Date[3] |
| --- | --- | --- |
| Bitcoin mixers | *cf.* Möser, Böhme & Breuker 2013 | 2011 |
| Bloom filter | Hearn & Corallo 2012 | October 2012 |
| CoinJoin | Maxwell 2013a | August 2013 |
| CoinShuffle | Ruffing 2014 | April 2014 |
| CoinSwap | Maxwell 2013b | October 2013 |
| Confidential transactions | Maxwell 2015 | June 2015 |
| Merge avoidance | Hearn 2013 | December 2013 |
| Mixcoin | Bonneau et al. 2014 | February 2014 |
| Ring signatures | Van Saberhagen 2012 | December 2012 |
| Stealth addresses | Todd 2014 | January 2014 |
| TumbleBit | Heilman et al. 2017 | June 2016 |
| XIM | Bissias et al. 2014 | November 2014 |
| Zerocash | Sasson et al. 2014 | May 2014 |
| Zerocoin | Miers et al. 2013 | May 2013 |

---

[3] The dates represent the earliest public proposals of the respective techniques that we could find. For the latest version of this table, see https://github.com/maltemoeser/bitcoin-anonymity/blob/master/obfuscation-techniques.md

**Appendix 2: The Lifecycle of Obfuscation**

The success of obfuscation in Bitcoin motivates studying the adoption of obfuscation in sociotechnical systems more generally. To this end, we present a simplified model of the adoption of obfuscation, visualized in Figure 2.

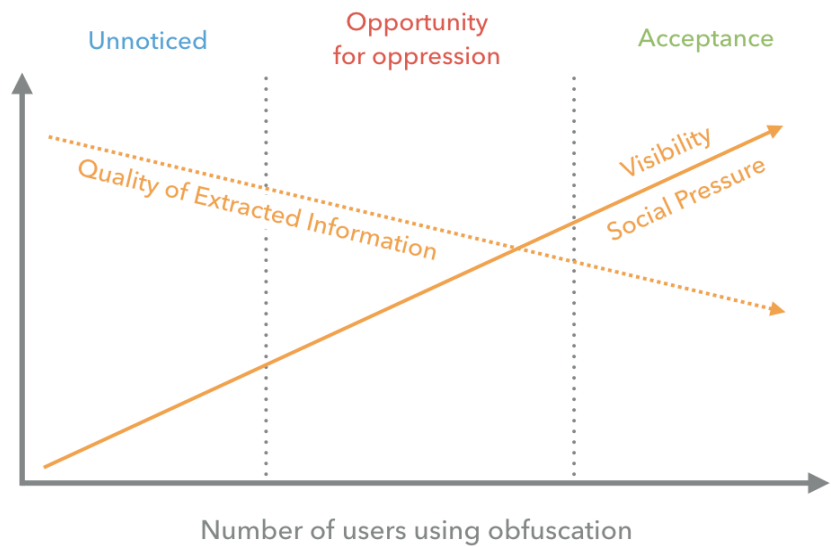

**Figure 2:** Lifecycle of Obfuscation

We conjecture that as the number of users of obfuscation grows, the visibility of the use of obfuscation increases as well. It also reduces the quality of the information that can be extracted from the system. We argue that initially, the use of obfuscation is mostly unnoticed as the user base and its impact is small. On the other hand, once obfuscation has reached a critical mass, social pressure helps against the platform owner's (or government's) wish to oppress obfuscation, leading to general acceptance. A good example is "Nymwars", i.e. Google's (and other companies) attempt to forbid the use of pseudonyms on social networks. Due to a large, negative public reaction Google had to reverse its decision to ban the use of pseudonyms. This suggests a critical phase in between these two, where there is opportunity for oppression by the platform owner or government. For those who aim to establish obfuscation as a means of defense against a system, this suggests two related strategies to minimize the window of oppression. The first is to hide the use of obfuscation for as long as possible through both social and technical means. The second is to maximize the visibility of obfuscation and campaign for its acceptance once it can no longer remain unnoticed.